\documentclass{elsart}
\def\intz{\int\limits_{z}^{1}\dd x\;}
\def\intzd{\int\limits_{z/(1-\Delta)}^{1-\Delta}\!\!\!\!\dd x\;}

\def\Ree{\mathop{\mathrm{Re}}}

\def\Ni#1#2{\mathop{\mathrm S\phantom{}}\nolimits_{#1}\left(#2\right)}
\def\Li#1#2{\mathop{\mathrm{Li}}\nolimits_{#1}\left(#2\right)}

\def\ba{\begin{eqnarray}}
\def\ea{\end{eqnarray}}
\def\dd{{\mathrm d}}

\def\fun#1#2{\lower3.6pt\vbox{\baselineskip0pt\lineskip.9pt
  \ialign{$\mathsurround=0pt#1\hfil##\hfil$\crcr#2\crcr\sim\crcr}}}

\begin{document}
\begin{frontmatter}

\title{Tables of convolution integrals}

\author{A.B. Arbuzov}

\address{Bogoliubov Laboratory for Theoretical Physics,\\
JINR, Dubna, 141980, Russia\\
{\tt e-mail:} arbuzov@thsun1.jinr.ru
}

\begin{abstract}
An analytical approach to convolution of functions, which appear
in perturbative calculations, is discussed. An extended list
of integrals is presented.
\end{abstract}


\end{frontmatter}

\section{Introduction}

There are many situations, where one can describe a certain
probability distribution of a complicated process in a form of
a conditional probability involving two or more sub--processes.
Typically in quantum physics, a factorization of
sub--processes occurs due to the presence of a small
(or large) parameter, which allows
to suppress the interference of amplitudes, describing different
sub--processes. We will consider the dependence on one continuous
variable $0\leq x \leq 1$, and assume that the conditional
probability can be presented as a convolution of the corresponding
distributions for the sub--processes.
In particular, convolution appear in the so--called evolution
equations, arising in the renormalization group approach.

Quite often the convolution is performed by using the Mellin
transformation. This approach is very transparent and powerful.
On the other hand, it requires more steps (direct and inverse
transformations) and involves a considerable number of auxiliary
functions in the moment space. Moreover, in a realistic application,
one might be interested to change the limits of convolution
integrals to separate a certain contribution with a particular physical
meaning.

Here I am going to discuss the direct analytical convolution, which is
known to work well with a rather wide class of functions, which appear
in perturbative calculations. The paper is organized as follows. In the
next section I introduce the notation. The tables of convolution integrals of
singular and non--singular functions are given in Sect.~\ref{intsg}
and Sect.~\ref{intng}, respectively. Possible applications of the Tables
are discussed in Conclusions. Properties of polylogarithmic functions
are sketched in Appendixes.

\section{Preliminaries}

Let us consider two functions $f(x)$ and $g(y)$, defined on the interval
$0 \leq x,y \leq 1$. Their convolution is given by
\ba
&& \bigl[f\otimes g\bigr](z) = \int\limits^1_0\dd x
\int\limits^1_0\dd y\; \delta(z-xy)f(x)g(y)
=\int\limits^1_z\frac{\dd x}{x}\; f(x)g\left(\frac{z}{x}\right),
\qquad 0 \leq z \leq 1.
\ea

Sometimes we need to perform a convolution of special functions.
In particular, functions with the so--called {\em plus prescription}
can be used. The prescription regularizes a pole singularity at $x=1$.
It is defined as follows:
\ba \label{plus}
&& \int\limits_{x_{\mathrm{min}}}^1\dd x\; [f(x)]_+ g(x) =
\int\limits_{0}^1\dd x\; f(x) [g(x)\Theta(x-x_{\mathrm{min}})-g(1)],
\\ \nonumber
&& \Theta(x) = \left\{\begin{array}{l} 1 \quad {\mathrm{for}} \quad
x\geq 0 \\ 0 \quad {\mathrm{for}} \quad x < 0
\end{array}\right. ,
\qquad 0 \leq x_{\mathrm{min}} < 1.
\ea

Integrals of functions, which have poles at $x=z$ or $x=1$, are
divergent. They can be regularized by introducing a small auxiliary
parameter $\Delta \ll 1$. In the final result of
a particular calculation,
one has than look for the cancellation of the parameter between
different contributions. The parameter can get also a physical meaning
as a separator between soft and hard radiation. There is a one--to--one
correspondence between the $\Delta$-regularization and the plus
prescription. In fact, the following definition is equivalent to
the prescription~(\ref{plus}):
\ba \label{delta}
&& [f(x)]_{+} = \lim_{\Delta\to 0}\bigl[ \delta(1-x)f_{\Delta}
+ \Theta(1-\Delta-x)f_{\Theta}(x)\bigr], \\ \nonumber
&& f_\Delta = - \int\limits_0^{1-\Delta}\dd x\; f(x), \qquad
f_{\Theta}(x) = f(x) \bigg|_{x<1}.
\ea
We will call $f_\Delta$ and $f_{\Theta}(x)$ as the $\delta$-part
and the $\Theta$-part of the special function $f(x)$.
The above realization of the plus prescription is convenient
in analytical calculations and especially in numerical computations,
where the direct implementation of Eq.~(\ref{plus}) is problematic.
In what follows we will systematically use $\Delta$-regularization
for divergent integrals, keeping in mind that the cancellation of
the parameter will happen after summing with the corresponding
contribution of the $\delta$-part of the relevant functions
as in Eq.~(\ref{delta}).

To define the $\delta$-part for a given function, one
requires a table of definite integrals over the interval $0<x<1-\Delta$
or $0<x<1$ for non--singular functions.
They can be found in numerous sources (see, for instance, Ref.~\cite{DD}).

A convolution of two singular functions regularized by the plus
prescription can be represented as
\ba
\biggl[[f]_{+}\otimes [g]_{+}\biggr](z) = \lim_{\Delta\to 0}\biggl\{
\int\limits^{1-\Delta}_{z/(1-\Delta)}\frac{\dd x}{x}
f_\Theta(x)g_\Theta \left(\frac{z}{x}\right)
+ f_\Delta g_\Theta(z) + f_\Theta(z) g_\Delta \biggr\}.
\ea

\section{Integrals of singular functions
\label{intsg}}

\ba
&& \intzd \frac{\ln^n(1-x)}{1-x} = \frac{1}{n+1}\ln^{n+1}(1-z)
- \frac{1}{n+1}\ln^{n+1}\Delta,
\qquad n = 0,1,2,\ldots \\
&& \intzd \frac{1}{x-z} = - \ln\Delta + \ln(1-z) - \ln z, \\
&& \intzd \frac{\ln x}{x-z} = - \ln\Delta \ln z + \Li{2}{1-z}
+ \ln(1-z)\ln z - \frac{1}{2}\ln^2z, \\
&& \intzd \frac{\ln(1-x)}{x-z} = - \ln\Delta \ln(1-z)
- \ln(1-z)\ln z + \ln^2(1-z) - \zeta(2), \\
&& \intzd \frac{\ln^2x}{x-z} = - \ln\Delta \ln^2z
+ 2\Ni{1,2}{1-z} + 2\Li{2}{1-z}\ln z
\nonumber \\ && \qquad
+ \ln(1-z)\ln^2z - \frac{1}{3}\ln^3z, \\
&& \intzd \frac{\ln^2(1-x)}{x-z} = - \ln\Delta \ln^2(1-z)
+ \ln^3(1-z) - \ln^2(1-z)\ln z
\nonumber \\ && \qquad
- 2\zeta(2)\ln(1-z) + 2\zeta(3), \\
&& \intzd \frac{\ln(1-x)\ln x}{x-z} = - \ln\Delta \ln(1-z)\ln z
+ 2\Ni{1,2}{1-z} - \Li{3}{1-z}
\nonumber \\ && \qquad
+ \Li{2}{1-z}\ln(1-z)
+ \Li{2}{1-z}\ln z + \ln^2(1-z)\ln z
\nonumber \\ && \qquad
- \frac{1}{2}\ln(1-z)\ln^2z
- \zeta(2)\ln z,\\
&& \intzd \frac{\Li{2}{1-x}}{x-z} = \Li{2}{1-z} \bigl( \ln(1-z)
- \ln z - \ln\Delta \bigr)
\nonumber \\ && \qquad
- \Ni{1,2}{1-z} - \Li{3}{1-z}, \\
&& \intzd \frac{\Li{3}{1-x}}{x-z} =
\Li{3}{1-z}\bigl(\ln(1-z) - \ln z - \ln\Delta\bigr)
\nonumber \\ && \qquad
- \Ni{2,2}{1-z} - \Li{4}{1-z}, \\
&& \intzd \frac{\Ni{1,2}{1-x}}{x-z} =
\Ni{1,2}{1-z}\bigl(\ln(1-z) - \ln z - \ln\Delta\bigr)
+ \Ni{2,2}{1-z}
\nonumber \\ && \qquad
- 2\Ni{1,3}{1-z} - \frac{1}{2}\left(\Li{2}{1-z}\right)^2,
\\
&& \intzd \frac{\Li{2}{1-x}\ln(1-x)}{x-z} =
\Li{2}{1-z}\ln(1-z)\bigl(\ln(1-z) - \ln z - \ln\Delta\bigr)
\nonumber \\ && \qquad
+ \Li{4}{1-z}
- 2\Ni{2,2}{1-z}
+ \frac{1}{2}\left(\Li{2}{1-z}\right)^2
- \Li{3}{1-z}\ln(1-z)
\nonumber \\ && \qquad
- \Ni{1,2}{1-z}\ln(1-z)
- \zeta(2)\Li{2}{1-z},
\\
&& \intzd \frac{\Li{2}{1-x}\ln x}{x-z} =
\Li{2}{1-z}\ln z\bigl(\ln(1-z) - \ln z - \ln\Delta\bigr)
\nonumber \\ && \qquad
+ 4\Ni{1,3}{1-z} - 4\Ni{2,2}{1-z}
+ \left(\Li{2}{1-z}\right)^2
- \Li{3}{1-z}\ln z
\nonumber \\ && \qquad
+ \Ni{1,2}{1-z}\ln z
+ \frac{1}{2}\Li{2}{1-z}\ln^2z,
\\
&& \intzd \frac{\ln^2(1-x)\ln x}{x-z} =
- \ln\Delta\ln^2(1-z)\ln z
+ 2\Li{4}{1-z}
\nonumber \\ && \qquad
- \left(\Li{2}{1-z}\right)^2
- 2\Li{3}{1-z}\bigl(\ln(1-z)+\ln z\bigr)
+ 4\Ni{1,2}{1-z}\ln(1-z)
\nonumber \\ && \qquad
+ 2\Li{2}{1-z}\ln(1-z)\ln z
+ \Li{2}{1-z}\ln^2(1-z)
+ \ln^3(1-z)\ln z
\nonumber \\ && \qquad
- \frac{1}{2}\ln^2(1-z)\ln^2z
- 2\zeta(2)\ln(1-z)\ln z
+ 2\zeta(3)\ln z,
\\
&& \intzd \frac{\ln(1-x)\ln^2x}{x-z} =
- \ln\Delta\ln(1-z)\ln^2z
+ 2\Ni{2,2}{1-z} - \left(\Li{2}{1-z}\right)^2
\nonumber \\ && \qquad
- 2\Li{3}{1-z}\ln z
+ 2\Ni{1,2}{1-z}\bigl(\ln(1-z)+\ln z\bigr)
\nonumber \\ && \qquad
+ 2\Li{2}{1-z}\ln(1-z)\ln z
+ \Li{2}{1-z}\ln^2z
+ \ln^2(1-z)\ln^2z
\nonumber \\ && \qquad
- \frac{1}{3}\ln(1-z)\ln^3z
- \zeta(2)\ln^2z,
\\
&& \intzd \frac{\ln^3(1-x)}{x-z} =
\ln^3(1-z)\bigl(\ln(1-z) - \ln z - \ln\Delta\bigr)
- 3\zeta(2)\ln^2(1-z)
\nonumber \\ && \qquad
+ 6\zeta(3)\ln(1-z) - 6\zeta(4),
\\
&& \intzd \frac{\ln^3x}{x-z} =
- \ln\Delta\ln^3z
+ 6\Ni{1,3}{1-z}
+ 6\Ni{1,2}{1-z}\ln z
\nonumber \\ && \qquad
+ 3\Li{2}{1-z}\ln^2z
+ \ln(1-z)\ln^3z
- \frac{1}{4}\ln^4z.
\ea
In the above integrals I omitted terms, which are vanishing in the limit
$\Delta\to 0$.

\section{Integrals of non--singular functions
\label{intng}}

\ba
&& \intz x^n = \frac{1}{n+1}\biggl( 1 - z^{n+1} \biggr), \qquad n \neq -1, \\
&& \intz \frac{\ln^nx}{x} = - \frac{1}{n+1}\ln^{n+1}z,
\qquad n = 0,1,2,\ldots \\
&& \intz \frac{\ln^n(1-x)}{x} = (-1)^nn!\;\bigl[ \zeta(n+1)
- \Ni{1,n}{z} \bigr],
\qquad n = 1,2,3,\ldots \\
&& \intz \frac{\ln(1-x)\ln x}{x} = \Ni{1,2}{1-z}
- \frac{1}{2}\ln(1-z)\ln^2z, \\
&& \intz x^n \ln x = - \frac{z^{n+1}}{n+1}\ln z
- \frac{1}{(n+1)^2}\biggl( 1 - z^{n+1} \biggr), \qquad n = 0,1,2,\ldots \\
&& \intz \frac{\ln x}{x^n} = \frac{1}{z^{n-1}(n-1)^2}\biggl( (n-1)\ln z
+ 1 - z^{n-1} \biggr), \qquad n = 2,3,4,\ldots \\
&& \intz \frac{\ln^nx}{1-x} = (-1)^n n!\; \Ni{1,n}{1-z},
\qquad n = 1,2,3,\ldots \\
&& \intz \frac{\ln(1-x)\ln x}{1-x} = \Li{3}{1-z} - \Li{2}{1-z}\ln(1-z), \\
&& \intz x^n\ln(1-x) = \frac{1-z^{n+1}}{n+1}\ln(1-z)
+ \frac{1}{n+1}\biggl( \sum\limits_{k=1}^{n+1}\frac{z^k}{k}
- \Ni{1}{n+1}\biggr),
\nonumber \\ && \qquad
n = 0,1,2,\ldots \\
&& \intz \frac{\ln(1-x)}{x^2} = \ln z + \frac{1-z}{z}\ln(1-z), \\
&& \intz \frac{\ln(1-x)}{x^n} = \frac{1}{n-1}\biggl( \ln z
+ \frac{1-z^{n-1}}{z^{n-1}}\ln(1-z)
- \sum\limits_{k=1}^{n-2}\frac{1}{kz^{k}} + \Ni{1}{n-2} \biggr),
\nonumber \\ && \qquad
n = 3,4,5,\ldots \\
&& \intz x^n\ln^2x = \frac{2}{(n+1)^3}(1-z^{n+1})
+ \frac{2z^{n+1}}{(n+1)^2}\ln z - \frac{z^{n+1}}{n+1}\ln^2z,
\quad n \neq -1,\\
&& \intz x^n\ln^2(1-x) = \frac{1-z^{n+1}}{n+1}\ln^2(1-z)
+ \sum\limits_{k=0}^{n}{n\choose{k}} (-1)^k
\frac{(1-z)^{k+1}}{(k+1)^2}\biggl( \frac{2}{k+1}
\nonumber \\ && \qquad
- 2\ln(1-z) \biggr), \qquad n = 0,1,2,\ldots \\
&& \intz x^n\ln(1-x)\ln x = - \frac{1}{n+1}\Li{2}{1-z}
- \frac{z^{n+1}}{n+1}\ln(1-z)\ln z
\nonumber \\ && \qquad
- \frac{1-z^{n+1}}{(n+1)^2}\ln(1-z)
+ \frac{1}{n+1}\ln z\sum\limits_{k=1}^{n+1}\frac{z^k}{k}
\nonumber \\ && \qquad
+ \frac{1}{(n+1)^2}\sum\limits_{k=1}^{n+1}(1-z^k)\frac{n+k+1}{k^2}\, ,
\qquad n=0,1,2,\ldots  \\
&& \intz \frac{\ln x\ln(1-x)}{x^n} = \frac{1}{n-1}\Li{2}{1-z}
+ \frac{1}{z^{n-1}(n-1)}\ln z\ln(1-z)
\nonumber \\ && \qquad
+ \ln z\biggl( \frac{1}{(n-1)^2}
- \frac{1}{n-1}\sum\limits_{k=1}^{n-2}\frac{1}{z^kk} \biggr)
+ \frac{1-z^{n-1}}{z^{n-1}(n-1)^2}\ln(1-z)
\nonumber \\ && \qquad
+ \frac{1}{2(n-1)}\ln^2z
- \frac{1}{(n-1)^2}\sum\limits_{k=1}^{n-2}
\frac{(1-z^k)(n+k-1)}{z^kk^2}\, ,
\quad n=2,3,4,\ldots  \\
&& \intz \frac{\ln^2(1-x)}{x^2} = \frac{1-z}{z}\ln^2(1-z)
+ 2\Li{2}{1-z} + 2\ln(1-z)\ln z, \\
&& \intz \frac{\ln^2(1-x)}{x^n} = \frac{1-z^{n-1}}{(n-1)z^{n-1}}\ln^2(1-z)
+ \frac{2}{n-1}\biggl( \Li{2}{1-z} + \ln(1-z)\ln z \biggr)
\nonumber \\ && \qquad
- \frac{2}{n-1}\sum\limits_{k=2}^{n-1}\intz \frac{\ln(1-x)}{x^k}\,
\qquad n = 3,4,5,\ldots
\ea

\ba
&& \intz \ln^2(1-x)\ln x = 2\Li{3}{1-z}
- 2\Li{2}{1-z}\ln(1-z)
\nonumber \\ && \qquad
- z\ln^2(1-z)\ln z
+ 2\Li{2}{1-z}
+ 2z\ln(1-z)\ln z
- (1-z)\ln^2(1-z)
\nonumber \\ && \qquad
+ 4(1-z)\ln(1-z)
- 2z\ln z + 6z - 6,
\\
&& \intz \frac{\ln^2(1-x)\ln x}{x} =
2\Ni{1,2}{1-z}\ln(1-z) - 2\Ni{2,2}{1-z}
- \frac{1}{2}\ln^2(1-z)\ln^2z,
\\
&& \intz \frac{\ln^2(1-x)\ln x}{x^2} =
- 2\Ni{1,2}{1-z} - 2\Li{3}{1-z} + 2\Li{2}{1-z}\ln(1-z)
\nonumber \\ && \qquad
+ \frac{1}{z}\ln^2(1-z)\ln z
+ \ln(1-z)\ln^2z
+ 2\Li{2}{1-z} + 2\ln(1-z)\ln z
\nonumber \\ && \qquad
+ \frac{1-z}{z}\ln^2(1-z),
\\
&& \intz \frac{\ln^2(1-x)\ln x}{1-x} =
- 2\Li{4}{1-z} + 2\Li{3}{1-z}\ln(1-z)
\nonumber \\ && \qquad
- \Li{2}{1-z}\ln^2(1-z),
\\
&& \intz \ln(1-x)\ln^2x = 2\Ni{1,2}{1-z}
- z\ln(1-z)\ln^2z
+ 2\Li{2}{1-z}
\nonumber \\ && \qquad
+ 2z\ln(1-z)\ln z + z\ln^2z + 2(1-z)\ln(1-z) - 4z\ln z
- 6 + 6z,
\\
&& \intz \frac{\ln(1-x)\ln^2x}{x} = - 2\Ni{1,3}{1-z}
- \frac{1}{3}\ln(1-z)\ln^3z,
\\
&& \intz \frac{\ln(1-x)\ln^2x}{x^2} = - 2\Ni{1,2}{1-z}
+ \frac{1}{z}\ln(1-z)\ln^2z + \frac{1}{3}\ln^3z
+ 2\Li{2}{1-z}
\nonumber \\ && \qquad
+ \frac{2}{z}\ln(1-z)\ln z
+ \ln^2z + 2\frac{1-z}{z}\ln(1-z) + 2\ln z,
\\
&& \intz \frac{\ln(1-x)\ln^2x}{1-x} =
- 2\Ni{2,2}{1-z} + 2\Ni{1,2}{1-z}\ln(1-z),
\\
&& \intz \ln^3(1-x) = (1-z)\bigl( \ln^3(1-z) - 3\ln^2(1-z)
+ 6\ln(1-z) - 6 \bigr),
\\
&& \intz \frac{\ln^3(1-x)}{x^2} = - 6\Li{3}{1-z}
+ 6\Li{2}{1-z}\ln(1-z) + 3\ln^2(1-z)\ln z
\nonumber \\ && \qquad
+ \frac{1-z}{z}\ln^3(1-z),
\\
&& \intz \ln^3x = - z\ln^3z + 3z\ln^2z - 6z\ln z + 6z - 6,
\\
&& \intz \frac{\ln^3x}{x^2} = \frac{1}{z}\bigl( \ln^3z
+ 3\ln^2z + 6\ln z - 6z + 6 \bigr).
\ea

By means of identical relations (see Appendix~B)
we reduce the arguments of the polylogarithm functions to $(1-x)$.
On the right hand side of the integrals we perform the same reduction
of arguments. It's worth to note, that there are certain physical
arguments in favor of the $(1-x)$ argument with respect to the simple
$x$. Namely, the point $x=1$ corresponds usually to a singularity
of a fragmentation or structure function, remind {\it e.g.} the
common lowest order splitting function
\ba
P^{(0)}(x) = \left[\frac{1+x^2}{1-x}\right]_+.
\ea
So in analytical results of a certain convolution, the limit $x\to1$
has usually a principal importance, and the choice of the argument
of polylogarithms helps to analyze this limit. Anyway, conversion
of functions depending on $(1-x)$ into the ones depending on $x$
can be performed using simple formulae from Appendix~B.
The relevant integrals of polylogarithmic functions are
\ba
&& \intz x^n\Li{2}{1-x} = \frac{1-z^{n+1}}{n+1}\Li{2}{1-z}
- \frac{1}{n+1}\sum\limits_{k=1}^{n+1}\biggl(
\frac{z^k}{k}\ln z + \frac{1-z^k}{k^2} \biggr),
\nonumber \\ && \qquad
n = 0,1,2,\ldots \\
&& \intz \frac{\Li{2}{1-x}}{1-x} = \Li{3}{1-z}, \\
&& \intz \frac{\Li{2}{1-x}}{x} = - 2\Ni{1,2}{1-z} - \Li{2}{1-z}\ln z, \\
&& \intz \frac{\Li{2}{1-x}}{x^n} = \frac{1-z^{n-1}}{z^{n-1}(n-1)}\Li{2}{1-z}
\nonumber \\ && \qquad
+ \frac{1}{n-1}\biggl[
\sum\limits_{k=1}^{n-2}\biggl( \frac{1-z^k}{z^kk^2}
+ \frac{\ln z}{z^kk} \biggr) - \frac{1}{2}\ln^2z\biggr],
\quad n = 2,3,4,\ldots
\ea

\ba
&& \intz \Li{3}{1-x} = (1-z)\Li{3}{1-z} - (1-z)\Li{2}{1-z} + z\ln z
+ 1 - z, \\
&& \intz \frac{\Li{3}{1-x}}{x} = - \Li{3}{1-z}\ln z
- \frac{1}{2}\left(\Li{2}{1-z}\right)^2,
\\
&& \intz \frac{\Li{3}{1-x}}{x^2} = \frac{1-z}{z}\Li{3}{1-z}
+ 2\Ni{1,2}{1-z} + \Li{2}{1-z}\ln z,
\\
&& \intz \frac{\Li{3}{1-x}}{1-x} = \Li{4}{1-z},
\\
&& \intz \Ni{1,2}{1-x} = (1-z)\Ni{1,2}{1-z}
+ \frac{z}{2}\ln^2z - z\ln z + z - 1,
\\
&& \intz \frac{\Ni{1,2}{1-x}}{x} = - 3\Ni{1,3}{1-z} - \Ni{1,2}{1-z}\ln z,
\\
&& \intz \frac{\Ni{1,2}{1-x}}{x^2} = \frac{1-z}{z}\Ni{1,2}{1-z}
+ \frac{1}{6}\ln^3z,
\\
&& \intz \frac{\Ni{1,2}{1-x}}{1-x} = \Ni{2,2}{1-z},
\\
&& \intz \Li{2}{1-x}\ln(1-x) = (1-z)\Li{2}{1-z}\ln(1-z)
+ (z-2)\Li{2}{1-z}
\nonumber \\ && \qquad
- z\ln(1-z)\ln z
- (1-z)\ln(1-z) + 2z\ln z
+ 3 - 3z,
\\
&& \intz \frac{\Li{2}{1-x}\ln(1-x)}{x} = 2\Ni{2,2}{1-z}
- \frac{1}{2}\left(\Li{2}{1-z}\right)^2
\nonumber \\ && \qquad
- 2\Ni{1,2}{1-z}\ln(1-z)
- \Li{2}{1-z}\ln(1-z)\ln z,
\\
&& \intz \frac{\Li{2}{1-x}\ln(1-x)}{x^2} = 3\Ni{1,2}{1-z}
+ \Li{2}{1-z}\ln z
\nonumber \\ && \qquad
+ \frac{1-z}{z}\Li{2}{1-z}\ln(1-z)
- \frac{1}{2}\ln(1-z)\ln^2z,
\\
&& \intz \frac{\Li{2}{1-x}\ln(1-x)}{1-x} =
- \Li{4}{1-z} + \Li{3}{1-z}\ln(1-z),
\\
&& \intz \Li{2}{1-x}\ln x = - 2\Ni{1,2}{1-z}
- (z\ln z - z + 1)\Li{2}{1-z} - z\ln^2z
\nonumber \\ && \qquad
+ 3z\ln z - 3z + 3,
\\
&& \intz \frac{\Li{2}{1-x}\ln x}{x} = 3\Ni{1,3}{1-z}
- \frac{1}{2}\Li{2}{1-z}\ln^2z,
\\
&& \intz \frac{\Li{2}{1-x}\ln x}{x^2} = 2\Ni{1,2}{1-z}
+ \frac{1}{z}(\ln z - z + 1)\Li{2}{1-z}
\nonumber \\ && \qquad
- \frac{1}{3}\ln^3z - \frac{1}{2}\ln^2z,
\\
&& \intz \frac{\Li{2}{1-x}\ln x}{1-x} =
- \frac{1}{2}\left(\Li{2}{1-z}\right)^2.
\ea

Integrals of some functions, which depend on $(1+x)$,
are required in certain cases (we consider only the real part
of the corresponding functions):
\ba
&& \intz \frac{1}{1+x} = \ln2 - \ln(1+z), \\
&& \intz \frac{\ln x}{1+x} = \Li{2}{1+z} - \frac{3}{2}\zeta(2), \\
&& \intz \frac{\ln^2x}{1+x} = \frac{7}{2}\zeta(3) - 2\Ni{1,2}{1+z}, \\
&& \intz \Li{2}{1+x} = - (1+z)\Li{2}{1+z} - z\ln z - 1 + z + 3\zeta(2), \\
&& \intz\frac{\Li{2}{1+x}}{1+x} = - \Li{3}{1+z} + \frac{7}{8}\zeta(3)
+ \frac{3}{2}\zeta(2)\ln2, \\
&& \intz\frac{\Li{2}{1+x}}{x} = \frac{7}{2}\zeta(3) - 2\Ni{1,2}{1+z}
- \Li{2}{1+z}\ln z, \\
&& \intz\frac{\Li{2}{1+x}}{x^n} = - \frac{3}{2(n-1)}\zeta(2)
+ \frac{1}{z^{n-1}(n-1)}\Li{2}{1+z}
\nonumber \\ && \qquad
- \frac{(-1)^{n-1}}{n-1}\biggl( \Li{2}{1+z} + \frac{1}{2}\ln^2z
- \frac{3}{2}\zeta(2) \biggr)
\nonumber \\ && \qquad
- \frac{1}{n-1}\sum\limits_{k=1}^{n-2}(-1)^{n+k}\biggl(
\frac{1-z^k}{z^kk^2} + \frac{\ln z}{z^kk} \biggr),
\qquad n = 2,3,4,\ldots
\ea

\section{Conclusions}

The tables of integrals were implemented in a
{\tt FORM}~\cite{Vermaseren:2000nd} subroutine and used to perform
analytical calculations of various convolutions in
Refs.~\cite{Arbuzov:1999cq,Arbuzov:2002pp,Arbuzov:2002cn,Arbuzov:2002rp}.
The subroutine can be used to construct an automated program
for convolution of a rather wide class of functions, which appear in
perturbative QED and QCD calculations. In particular, a possibility to
include the leading and next--to--leading logarithmic corrections
into the SANC project~\cite{Andonov:2002jg} is considered. It can be done
by means of an automated convolution of perturbative coefficient functions with relevant structure and fragmentation functions.

Most of the presented integrals can be found in other sources,
including automatic integrators in {\tt MATHEMATICA}~\cite{math}
and other packages. But I hope that the tables can come in handy
in further analytical calculations.

\ack
This work is supported by the RFBR grant 03-02-17077.

\appendix
\section*{Appendix A\\
Notation for polylogarithm and other functions}

\setcounter{equation}{0}
\renewcommand{\theequation}{A.\arabic{equation}}

The short notation for partial sums is
\ba
\Ni{k}{n} = \sum\limits_{j=1}^{n}\frac{1}{j^k}\, ,
\qquad \Ni{k}{0} = 0.
\ea

Binomial coefficients are
\ba
{n\choose{k}} = \frac{n!}{k!(n-k)!}\, .
\ea

The Riemann $\zeta$--functions are defined as usual:
\begin{eqnarray}
\zeta(n) &=& \sum\limits_{k=1}^{\infty}\frac{1}{k^n}, \qquad
\zeta(2) = \frac{\pi^2}{6}\, , \qquad
\zeta(3) \approx 1.20205690315959, \nonumber \\
\zeta(4) &=& \frac{\pi^4}{90}\, , \qquad
\zeta(5) \approx 1.03692775514337\; .
\end{eqnarray}

Following the notation of Refs.~\cite{KMR,DD},
the general Nielsen's polylogarithm is
\begin{eqnarray}
&& \Ni{n,m}{z} = \frac{(-1)^{n+m-1}}{(n-1)!m!}
\int\limits_{0}^{1}\dd x\frac{\ln^{n-1}(x)\ln^m(1-xz)}{x},
\nonumber \\ && \qquad
n= 1,2,3\ldots, \qquad m= 1,2,3\ldots
\end{eqnarray}
In particular,
\begin{eqnarray}
\Li{2}{z} &\equiv& \Ni{1,1}{z} = - \int\limits_{0}^{1}
\dd x\frac{\ln(1-xz)}{x}\, , \qquad
\Ni{1,2}{z} = \frac{1}{2}\int\limits_{0}^{1}\dd x\frac{\ln^2(1-xz)}{x}\, ,
\nonumber \\
\Li{3}{z} &\equiv& \Ni{2,1}{z}
= \int\limits_{0}^{1}\dd x\frac{\ln(x)\ln(1-xz)}{x}
= \int\limits_{0}^{z}\dd x\frac{\Li{2}{x}}{x}\, ,
\nonumber \\
\Li{4}{z} &\equiv& \Ni{3,1}{z}
= - \frac{1}{2} \int\limits_{0}^{1}\dd x\frac{\ln^2(x)\ln(1-xz)}{x}\, ,
\qquad
\Ni{1,3}{z}
= - \frac{1}{6} \int\limits_{0}^{1}\dd x\frac{\ln^3(1-xz)}{x}\, ,
\nonumber \\
\Ni{2,2}{z} &=& - \frac{1}{2} \int\limits_{0}^{1}
\dd x\frac{\ln(x)\ln^2(1-xz)}{x}\, .
\end{eqnarray}

\section*{Appendix B\\
Relations between polylogarithms}

\setcounter{equation}{0}
\renewcommand{\theequation}{B.\arabic{equation}}

To convert polylogarithms of $z$ into the ones of $(1-z)$, one can use
the following relations:
\ba
&& \Li{2}{z} = - \Li{2}{1-z} - \ln z\ln(1-z) + \zeta(2),\nonumber  \\
&& \Ni{1,2}{z} = - \Li{3}{1-z} + \Li{2}{1-z}\ln(1-z)
+ \frac{1}{2}\ln^2(1-z)\ln z + \zeta(3),\nonumber  \\
&& \Li{3}{z} = - \Ni{1,2}{1-z} - \Li{2}{1-z}\ln z
- \frac{1}{2}\ln(1-z)\ln^2z + \zeta(2)\ln z + \zeta(3),
\nonumber \\
&& \Ni{1,3}{z} = - \Li{4}{1-z} + \Li{3}{1-z}\ln(1-z)
- \frac{1}{2}\Li{2}{1-z}\ln^2(1-z)
\nonumber \\ && \qquad
- \frac{1}{6}\ln^3(1-z)\ln z
+ \zeta(4),
\nonumber \\
&& \Ni{2,2}{z} = - \Ni{2,2}{1-z} - \Li{3}{1-z}\ln z
+ \Ni{1,2}{1-z}\ln(1-z)
\nonumber \\ && \qquad
+ \Li{2}{1-z}\ln(1-z)\ln z
+ \frac{1}{4}\ln^2(1-z)\ln^2z + \zeta(3)\ln z +\frac{1}{4}\zeta(4),
\nonumber \\
&& \Li{4}{z} = - \Ni{1,3}{1-z} - \Ni{1,2}{1-z}\ln z
- \frac{1}{2}\Li{2}{1-z}\ln^2z
\nonumber \\ && \qquad
- \frac{1}{6}\ln(1-z)\ln^3z
+ \frac{1}{2}\zeta(2)\ln^2z + \zeta(3)\ln z + \zeta(4).
\ea

Some explicit expressions for polylogarithms of constant arguments
can be useful:
\ba
&&  \Ni{n,m}{0} = 0, \qquad  n= 1,2,3\ldots, \quad m= 1,2,3\ldots
\nonumber \\
&& \Li{n}{1} = \zeta(n), \qquad n = 2,3,4,\ldots
\nonumber \\
&& \Ni{2,2}{1} = \frac{1}{4}\zeta(4), \qquad
\Ni{1,n}{1} = \zeta(n+1), \qquad n= 1,2,3\ldots
\nonumber \\
&& \Ree\Li{2}{2} = \frac{3}{2}\zeta(2), \qquad
\Li{2}{-1} = - \frac{1}{2}\zeta(2), \qquad
\Li{2}{\frac{1}{2}} = \frac{1}{2}\zeta(2) - \frac{1}{2}\ln^22,
\nonumber \\
&& \Ree\Li{3}{2} = \frac{7}{8}\zeta(3) + \frac{3}{2}\ln 2\zeta(2),
\qquad
\Ree\Li{3}{-1} = - \frac{3}{4}\zeta(3), \qquad
\nonumber \\
&& \Li{3}{\frac{1}{2}} = \frac{7}{8}\zeta(3) - \frac{1}{2}\zeta(2)\ln2
+ \frac{1}{6}\ln^32,
\nonumber \\
&& \Ree\Ni{1,2}{2} = \frac{7}{4}\zeta(3), \qquad
\Ni{1,2}{-1} = \frac{1}{8}\zeta(3), \qquad
\Ni{1,2}{\frac{1}{2}} = \frac{1}{8}\zeta(3) - \frac{1}{6}\ln^32.
\ea

\end{document}